\documentclass[aps,prl,twocolumn,superscriptaddress]{revtex4-1}

\usepackage[dvipdf]{graphicx}

\usepackage{amssymb}

\usepackage{amsbsy}

\usepackage{latexsym}

\begin{document}

\title{Strain-dependent Splitting of Double Resonance Raman Scattering Band in Graphene}

\author{Duhee \surname{Yoon}}
\affiliation{Department of Physics, Sogang University, Seoul 121-742, Korea}

\author{Young-Woo \surname{Son}}
\email{hand@kias.re.kr}
\affiliation{School of Computational Sciences, Korea Institute for Advanced Study, Seoul 130-722, Korea}

\author{Hyeonsik \surname{Cheong}}
\email{hcheong@sogang.ac.kr}
\affiliation{Department of Physics, Sogang University, Seoul 121-742, Korea}

\date{28 Decemebr 2010}

\begin{abstract}
Under homogeneous uniaxial strains, the Raman 2$D$ band of graphene involving two-phonon double-resonance scattering processes splits into two peaks and they altogether redshift strongly depending on the direction and magnitude of the strain. Through polarized micro- Raman measurements and first-principles calculations, the effects are shown to originate from significant changes in resonant conditions owing to both the distorted Dirac cones and anisotropic modifications of phonon dispersion under uniaxial strains. Quantitative agreements between the calculation and experiment enable us to determine the dominant double- resonance Raman scattering path, thereby answering a fundamental question concerning this key experimental analyzing tool for graphitic systems.
\begin{description}
\item[PACS numbers]
63.22.Rc, 63.20.dk, 73.22.Pr.
\end{description}
\end{abstract}

\maketitle

The effects of external mechanical perturbations on physical properties of graphene are attracting much attention because of the possible realization of synthetic electromagnetic fields \cite{Vozmediano,Pereira,Choi1,Guinea,Choi2,Levy} and determinations of its fundamental material parameters \cite{Mohiuddin,Huang1,Frank,Ding,Mohr1,Proctor,Tsoukleri,Gong}. Since electronic structures and phonon dispersion are modified under applied strains \cite{Vozmediano,Pereira,Choi1,Choi2,Levy,Mohiuddin,Huang1,Frank,Ding,Mohr1,Proctor,Tsoukleri,Gong,Mohr2,Huang2}, the Raman spectrum -- an important diagnostic tool for graphitic systems -- will show significant variations. Recent experiments demonstrate that the Raman $G$ band redshifts and splits into two peaks under strain because of symmetry breaking of the doubly degenerate $E_{2g}$ phonons \cite{Mohiuddin,Huang1,Frank}. Furthermore, one can determine the Gr\"{u}neisen parameter of graphene \cite{Mohiuddin,Huang1,Ding} and identify its crystallographic orientation \cite{Mohiuddin,Huang1}.

Unlike the Raman $G$ band, the scattering process of the Raman 2$D$ band involves electronic states and TO phonons near the $K$ and $K'$ points of the Brillouin zone so that the strain-induced anisotropy of the electronic band structure \cite{Pereira,Choi1} and the phonon dispersion \cite{Mohr1,Mohr2} must manifest themselves in the Raman spectra of strained graphene. Although there have been experimental analyses assuming isotropic TO phonon softening \cite{Huang2} and independent theoretical studies regarding the 2$D$ band of strained graphene \cite{Mohr2}, a comprehensive and systematic study considering changes in both electronic and phonon structures is still lacking. Moreover, because several resonant scattering processes contribute to the 2$D$ band, a fundamental question concerning the dominant double-resonance process remains to be resolved \cite{Mafra1}.

In this Letter, we present a comprehensive analysis of the changes in electronic energy bands and phonon dispersion of a single-layer graphene under homogeneous uniaxial strains by combining polarized Raman measurements with an analysis based on first-principles calculations and determine the dominant scattering path of the double-resonance Raman scattering process. As the magnitude of the strain increases, the Raman 2$D$ band is split into two peaks, both of which redshift. Moreover, two distinct strains applied along armchair and zigzag crystallographic directions are identified and the frequency shift rate for each split Raman 2$D$ peak is strongly dependent on the strain direction. From theoretical analysis, we demonstrate that the anisotropic TO phonon softening together with distortions of Dirac cones is a dominant factor responsible for the observed effects. Furthermore, the polarization dependence of the relative intensities of the split 2$D$ band components reveals contributions of different resonant scattering paths, thereby establishing fundamental understanding of the double-resonance Raman scattering process in graphene.

Single-layer graphene samples were prepared on acrylic substrates with $50\times10\times1.3~ \textrm{mm}^{3}$ dimensions by using the micromechanical cleaving method from natural graphite flakes \cite{Novoselov}. Single-layer graphene samples were identified with micro- Raman spectroscopy \cite{Ferrari,Graf,Yoon1}. The results from two samples with special orientations will be compared. Strain was applied by bending the substrate with a specially designed jig. The Raman spectra were obtained using a polarized micro-Raman system using the 514.5-nm line of an Ar ion laser as the excitation. Other experimental details have been previously published \cite{Yoon2}.

\begin{figure} [b!]
\includegraphics{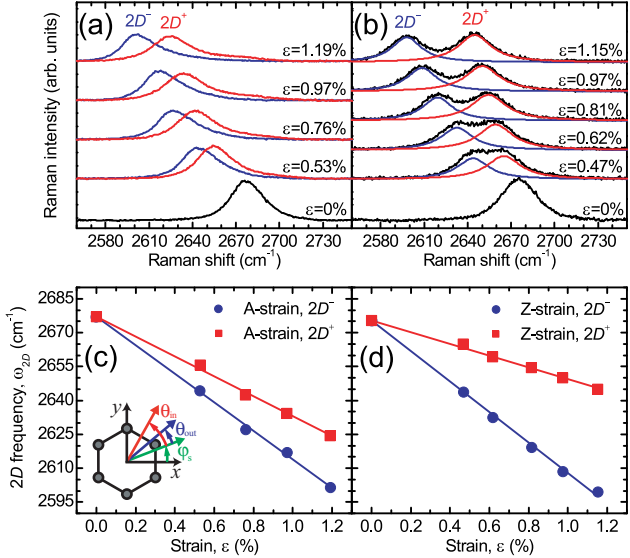}
\caption{(Color online) Evolutions of the 2$D$ bands of (a) A-strain and (b) Z-strain samples as a function of uniaxial strain. In (a), the 2$D^{-}$  and 2$D^{+}$  peaks are obtained with  $\theta_{\textrm{in}}=\theta_{\textrm{out}}=0^{\circ}$ and  $\theta_{\textrm{in}}=\theta_{\textrm{out}}=90^{\circ}$, respectively. In (b), the spectra are measured with $\theta_{\textrm{in}}=\theta_{\textrm{out}}=50^{\circ}$ and deconvoluted into two Lorentzian peaks. The positions of the 2$D^{-}$ and 2$D^{+}$ peaks of (c) the A-strain and (d) Z-strain samples as a function of strain. The solid lines are linear fits to the data. The inset shows the polarization geometry, where $\theta_{\textrm{in}}$, $\theta_{\textrm{out}}$, and $\varphi_{S}$ are the angles that the incident laser polarization, the analyzer axis, and the zigzag direction make with respect to the strain axis, respectively.}
\label{fig:fig1.eps}
\end{figure}

As the magnitude of uniaxial strain ($\epsilon$) increases, the $G$ band redshifts and splits into two peaks, $G^{-}$ and $G^{+}$ (not shown), as was reported earlier \cite{Mohiuddin,Huang1,Frank}. The shift rates are $\partial\omega_{G^{-}}/\partial\epsilon=-33.4~\textrm{cm}^{-1}/\%~(-33.0~\textrm{cm}^{-1}/\%)$ and $\partial\omega_{G^{+}}/\partial\epsilon=-14.5~\textrm{cm}^{-1}/\% ~(-12.9~\textrm{cm}^{-1}/\%)$ for sample \textit{A} (\textit{B}). They are essentially the same for the two samples as expected for small strain from symmetry and are in agreement with Ref.~7 but larger than the values in Ref.~8, probably due to difference in strain calibration. We also obtain the Gr\"{u}neisen parameter of $2.2\pm0.1$ and the shear deformation potential of $0.93\pm0.04$, in excellent agreement with previous estimations \cite{Mohiuddin,Thomsen1}. The angle between the strain direction and the zigzag direction of the graphene lattice, $\varphi_{S}$, was found to be $34.9\pm0.2^{\circ}$ $ (52.7\pm0.5^{\circ}$) for sample \textit{A} (\textit{B}) from the polarization dependence of the relative intensities of the $G^{-}$ and $G^{+}$ peaks \cite{Mohiuddin,Huang1}. Since $\varphi_{S}=30^{\circ}$ and $60^{\circ}$ for strain applied exactly along the armchair and zigzag directions, we will henceforth refer to samples \textit{A} and \textit{B} as A-strain and Z-strain samples, respectively.

When strain is applied, the 2$D$ band splits into two peaks which redshift as the strain increases [Figs.~\ref{fig:fig1.eps}(a) and (b)]. Unlike the $G$ band, the frequency shift rates of the 2$D^{-}$ and 2$D^{+}$ peaks of the A- and Z-strain samples are significantly different from each other: for A strain, $\partial\omega_{2D^{-}}/\partial\epsilon=-63.1~\textrm{cm}^{-1}/\%$ and $\partial\omega_{2D^{+}}/\partial\epsilon=-44.1~\textrm{cm}^{-1}/\%$, whereas for Z strain, $\partial\omega_{2D^{-}}/\partial\epsilon=-67.8~\textrm{cm}^{-1}/\%$ and $\partial\omega_{2D^{+}}/\partial\epsilon=-26.0~\textrm{cm}^{-1}/\%$. Again, these values are larger than those in Ref.~16, presumably due to difference in strain calibration.

\begin{figure} [b!]
\includegraphics{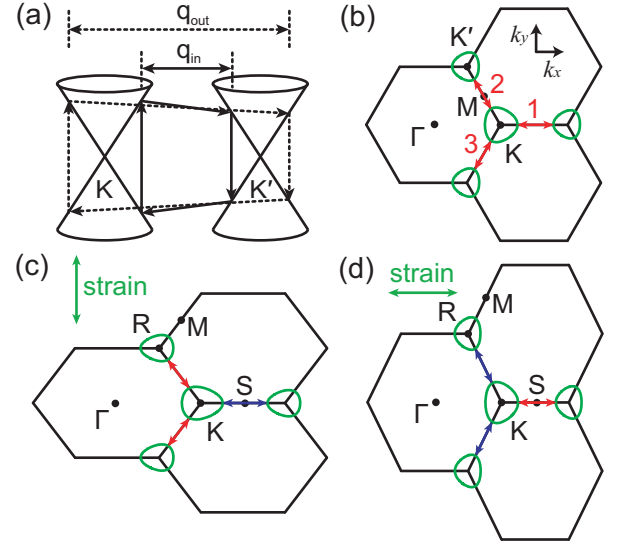}
\caption{(Color) (a) Double-resonance Raman scattering process. Inner ($q_{\textrm{in}}$) and outer ($q_{\textrm{out}}$) processes are indicated. Reciprocal lattice diagram for (b) unstrained, (c) A-strained, and (d) Z-strained graphene, showing strain-induced distortions. The $K'$ points are now designated as $R$ points, and there are inequivalent high symmetry points $M$ and $S$, the midpoint between the $K$ and $R$ points.}
\label{fig:fig2.eps}
\end{figure}

The dependence of strain-induced 2$D$ band splitting on the strain direction offers a unique opportunity to examine the strain-induced anisotropy of the electronic and phonon bands. The 2$D$ band comes from the four-step Stokes-Stokes double-resonance Raman scattering as illustrated in Fig.~\ref{fig:fig2.eps}(a) \cite{Ferrari,Yoon2,Thomsen2,Malard}. Theoretical calculations \cite{Maultzsch,Mafra2} suggested that the scattering processes involving the smallest momentum transfer (inner process) and the largest momentum transfer (outer process) are dominant contributions, but it is still not clear which of the two is the dominant one \cite{Mafra1}. For a given laser wavelength, the momentum of the emitted phonon is determined by the electronic band structure and the phonon dispersion near the $K$ and $K'$ points. For unstrained graphene, the scattering processes involving the three $K'$ points around a given $K$ point [denoted by 1, 2 and 3 in Fig.~ \ref{fig:fig2.eps}(b)] are completely equivalent, and hence the 2$D$ band appears as a single peak. When A strain is applied, the reciprocal lattice is distorted, as in Fig.~\ref{fig:fig2.eps}(c); one of the three $K'$ points moves away whereas the other two $K'$ points move closer to the $K$ point. Therefore, the two types of scattering processes involve phonons with different momenta, resulting in a splitting of the 2$D$ band. For Z strain, the distortion of the reciprocal lattice is reversed as in Fig.~\ref{fig:fig2.eps}(d). In Figs.~\ref{fig:fig2.eps}(c) and (d), the scattering process 1 involves phonons with momenta in the  $\mathit{\Gamma}KS$ direction, whereas processes 2 and 3 involve phonons with momenta in the $\mathit{\Gamma}RM$ direction.

In order to analyze the observed splitting of the 2$D$ band quantitatively, modifications of both the electronic band structure and the phonon dispersion due to strain must be taken into account. Strain shifts Dirac points away from the $K$ or $R$ points and tilts and distorts the Dirac cone so that the group velocity depends on the direction in the Brillouin zone \cite{Pereira,Choi1}. Strain also modifies the phonon dispersion. The observed splitting and softening of the 2$D$ band thus result from a convolution of the electronic band structure and the phonon dispersion modifications. So, we performed calculations on electronic band structures of strained single-layer graphene based on the first-principles self-consistent pseudopotential method \cite{Giannozzi} using the generalized gradient approximation for exchange-correlation functional \cite{Perdew} and on their phonon dispersions by using density-functional perturbation theory \cite{Giannozzi,Baroni}. The ion core of carbon atoms is described by an ultrasoft pseudopotential \cite{Vanderbilt}. A $k$-point sampling of $48\times48\times1$ grid uniformly distributed in the two-dimensional Brillouin zone is used in self-consistent calculations and a $6\times6\times1$ grid is used to calculate the dynamical matrices. The obtained electronic structures and phonon dispersions for graphene without strain are in good agreement with other studies [optical phonon frequencies at $\mathit{\Gamma}$ and $K$ points are $1581~\textrm{cm}^{-1}$ ($E_{2g}$) and $1295~\textrm{cm}^{-1}$ ($A'_1$), respectively.] \cite{Yan}.

\begin{figure}
\includegraphics{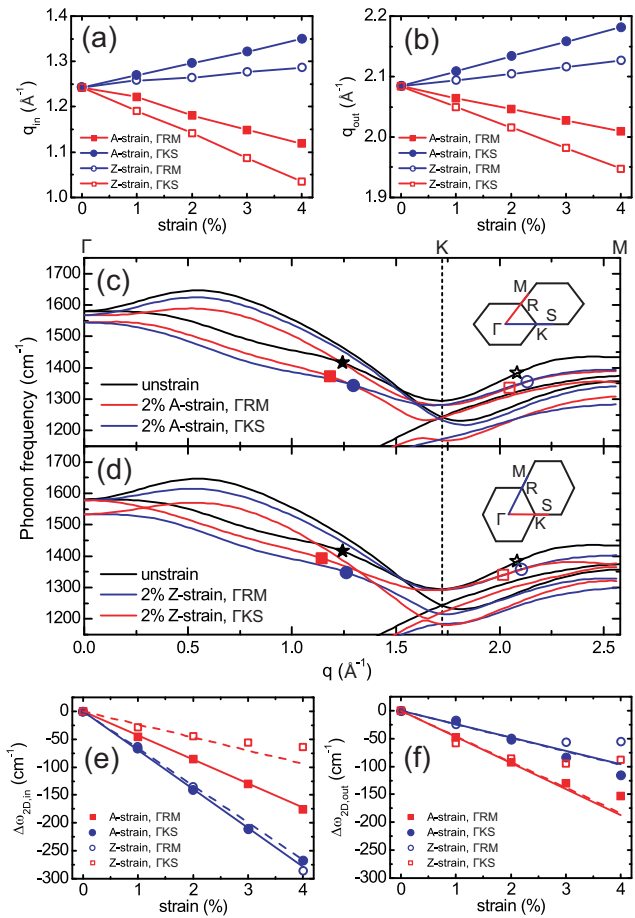}
\caption{(Color) Calculated phonon momenta involved in the (a) inner ($q_{\textrm{in}}$) and (b) outer ($q_{\textrm{out}}$) processes for the scattering in the $\mathit{\Gamma}KS$ and $\mathit{\Gamma}RM$ directions and A and Z strains. Phonon dispersions near the $K$ or $R$ points for 2\% strain applied in the (c) armchair and (d) zigzag direction. Filled and open stars are for the 2$D$ band in unstrained graphene, filled and open squares for the 2$D^{+}$ peak, and filled and open circles for the 2$D^{-}$ peak. Calculated strain dependences of the 2$D$ peaks for (e) inner and (f) outer processes.}
\label{fig:fig3.eps}
\end{figure}

The calculations show that the dominant contribution to the observed splitting and strain-direction-dependent frequency shifts originates from anisotropic changes of TO phonon branches with distorted Dirac cones as presented below. Since modifications of the electronic structure will change the resonant conditions as discussed above, the scattered phonon momentum will change their magnitude depending on their directions that are determined by momentum conservations under the strain. Figures \ref{fig:fig3.eps}(a) and (b) are the calculated phonon momenta satisfying the resonant conditions for the inner ($q_{\textrm{in}}$) and outer ($q_{\textrm{out}}$) processes, respectively. They are significantly different for the $\mathit{\Gamma}KS$ and $\mathit{\Gamma}RM$ directions and for A and Z strains. Figures \ref{fig:fig3.eps}(c) and (d) illustrate the modified phonon dispersions near the $K$ or $R$ points for $2\%$ strain applied in the armchair and zigzag directions, respectively. In the figure, the corresponding resonant frequencies for the Raman 2$D$ band are indicated by filled (inner) and open (outer) symbols, respectively. It should be noted that the phonon dispersions along the $\mathit{\Gamma}KS$ and $\mathit{\Gamma}RM$ directions are significantly different, especially away from the K point. This is in direct contradiction to the assumption used in Ref.~16 that the phonon softening rate is orientation independent. Figures \ref{fig:fig3.eps}(e) and (f) summarize the strain dependences of the 2$D$ peaks for the inner and outer processes, respectively. It is clear that the outer process is not consistent with our experimental data shown in Fig.~\ref{fig:fig1.eps}. A linear fit to the calculated values up to $2\%$ of strain gives, for the inner process [Fig.~\ref{fig:fig3.eps}(e)],
$\partial\omega_{2D^{-}}/\partial\epsilon=-70 \textrm{cm}^{-1}/\%$ and $\partial\omega_{2D^{+}}/\partial\epsilon=-43 \textrm{cm}^{-1}/\%$ for A strain and $\partial\omega_{2D^{-}}/\partial\epsilon=-66 \textrm{cm}^{-1}/\%$ and $\partial\omega_{2D^{+}}/\partial\epsilon=-24 \textrm{cm}^{-1}/\%$ for Z strain, in excellent quantitative agreement with the experimental data. Therefore, our Raman data for strained graphene clearly demonstrate that the inner process is the dominant one in the double- resonance Raman scattering. We also note that the softening of the TO phonon is more or less isotropic at high symmetric points ($K$ and $R$) as assumed in Ref.~16, but the strain phonon momentum satisfying double-resonance conditions deviates from those points where the effects of anisotropic softening are significant. It should be noted that the 2$D$ splitting rates for A and Z strain could not be explained in Ref.~16 with a model that assumed an isotropic TO phonon softening.

\begin{figure}
\includegraphics{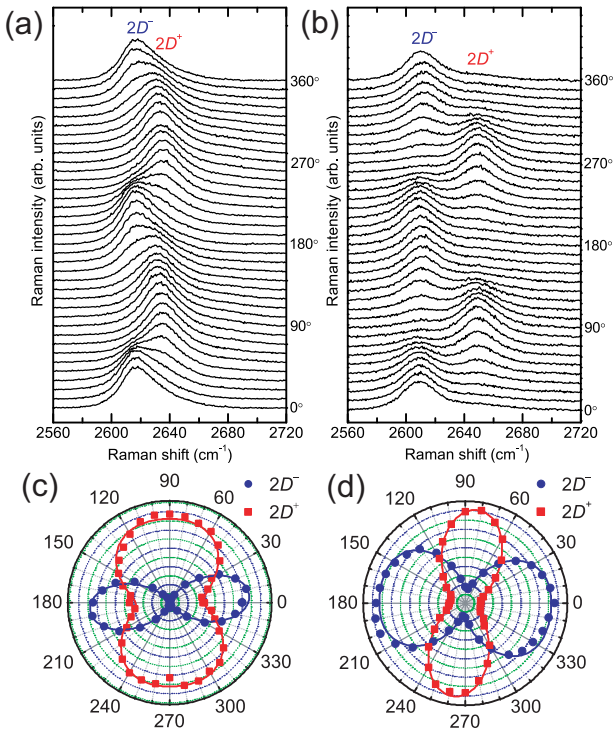}
\caption{(Color online) Evolutions of the 2$D$ band of the (a) A-strain and (b) Z-strain samples as a function of the incident laser polarization relative to the strain axis ($\theta_{\textrm{in}}$), under 0.97\% strain. The analyzer is parallel to the incident laser polarization. Polar plots of the 2$D^{-}$  and 2$D^{+}$ bands of the (c) A-strain and (d) Z-strain samples as a function of $\theta_{\textrm{in}}$.}
\label{fig:fig4.eps}
\end{figure}

Finally, the intensities of the 2$D^{-}$ and 2$D^{+}$ peaks depend strongly on the polarization direction of the incident laser [Figs.~\ref{fig:fig4.eps}(a) and (b)], corroborating our analysis. The intensities of the 2$D^{-}$ and 2$D^{+}$  peaks were measured as a function of the incident polarization angle ($\theta_{\textrm{in}}$) in steps of $10^{\circ}$. The analyzer is kept parallel to the incident polarization, $\theta_\textrm{in}=\theta_\textrm{out}$, which preferentially selects phonons in the direction orthogonal to $\theta_\textrm{in}$ \cite{Yoon2,Gruneis}. In A-strained graphene, for example, the phonons involved in the scattering processes of 2 and 3 in Fig.~\ref{fig:fig2.eps}(c) have the same frequency, whereas the phonons for the process of 1 differ. Since each of the three scattering processes contributes to the 2$D$ band, the peak corresponding to the process of 2 and 3 should have a contribution twice that of 1. This is demonstrated in Figs.~\ref{fig:fig4.eps}(c) and (d). In Fig.~\ref{fig:fig4.eps}(c), the 2$D^{-}$  band has one sinusoidal component fitted well to $I_{2D^{-}}\varpropto\cos^{4}(\theta_{\textrm{in}}-\phi_{1})$, whereas the 2$D^{+}$ band is fitted to two sinusoidal components of $I_{2D^{+}}\varpropto\cos^{4}(\theta_{\textrm{in}}-\phi_{2}-2\pi/3)+\cos^{4}(\theta_{\textrm{in}}-\phi_{2}-4\pi/3)$, where $\phi_{1}=5.1^{\circ}$ and $\phi_{2}=3.1^{\circ}$ \cite{Comment}. In Fig.~\ref{fig:fig4.eps}(d), the 2$D^{-}$ band has two sinusoidal components fitted well to $I_{2D^{-}}\varpropto\cos^{4}(\theta_{\textrm{in}}-\phi_{1}-7\pi/6)+\cos^{4}(\theta_{\textrm{in}}-\phi_{1}-11\pi/6)$, whereas the 2$D^{+}$  band, with one sinusoidal component, is fitted to $I_{2D^{+}}\varpropto\cos^{4}(\theta_{\textrm{in}}-\phi_{2}-\pi/2)$, where  $\phi_{1}=-6.0^{\circ}$ and $\phi_{2}=-7.5^{\circ}$.

In conclusion, the strain-induced splitting and redshift of the Raman 2$D$ band are found to depend on the direction of the applied strain with respect to crystallographic orientation. Comparison of experimental data with first-principles calculations shows that anisotropic modifications of the phonon dispersion together with changes in electronic structures are their origins. Furthermore, the dominant inner scattering process is demonstrated to resolve a controversy regarding the nature of the Raman 2$D$ band.

\emph{Note added}: Recently, we became aware of the publication of related work on the Raman 2$D$ band of strained graphene \cite{Huang2}, which was reported simultaneously with ours recently \cite{APS}.

\begin{acknowledgments}
We thank KIAS for providing computing resources (KIAS CAC Linux Cluster System). This work was supported by the National Research Foundation of Korean government (MEST) (No. 2008-0059038 and KRF-2008-314-C00111). D.~Y.~acknowledges funding from the Seoul City Government. Y.-W. S. was supported in part by the NRF grant funded by MEST (Quantum Metamaterials Research Center, R11-2008-053-01002-0 and Nano R$\&$D program 2008-03670).
\end{acknowledgments}


\begin{thebibliography}{99}
\bibitem{Comment}The 2$D$ Raman intensity for given $K'$ valley is proportional to the absorption and emission, which are proportion to $\textrm{cos}^2\theta_\textrm{in}$ and $\textrm{cos}^2\theta_\textrm{out}$, respectively. Since $\theta_\textrm{in}=\theta_\textrm{out}$, the Raman intensity for each $K'$ valley is proportional to $\textrm{cos}^4\theta_\textrm{in}$.
\end{thebibliography}
\end{document}